\documentclass[runningheads]{svmult}

\usepackage{makeidx}   
\usepackage{graphicx}  
\usepackage{subeqnar}  
\usepackage{multicol}  
\usepackage{physprbb}  
\makeindex             

\newcommand{\beq}{\begin{equation}}
\newcommand{\eeq}{\end{equation}}

\begin{document}

\title*{Guide for atomic and
particle physicists\protect\newline to CODATA's recommended
values\protect\newline of the fundamental physical constants }
\toctitle{Guide for atomic and particle physicists into CODATA's
recommended values of the fundamental physical constants
}
%
%
\titlerunning{Guide to CODATA's recommended values}
%
\author{Savely G. Karshenboim}
\authorrunning{Savely G. Karshenboim}
%
%
\institute{D. I. Mendeleev Institute for Metrology (VNIIM), St.
Petersburg 190005, Russia\protect\newline Max-Planck-Institut
f\"ur Quantenoptik, 85748 Garching, Germany}

\maketitle              

\begin{abstract}
The CODATA recommended values of the fundamental constants are
widely applied in particle, nuclear and atomic physics. They are a
result of a complicated evaluation (adjustment) of numerous
correlated data of different nature. Their application is often
rather mechanical and as a result is not free of various
confusions which are discussed in this note.
\end{abstract}

\def\lbar{\lambda\hskip-4.5pt\vrule height4.6pt depth-4.3pt width4pt}

\section{Introduction}

Precision physics deals with numbers rather than with functions,
but any theoretical prediction in numerical terms can appear only
after one applies certain values of input parameters, the most
important of which are the fundamental physical constants. The
most popular are recommended values published by CODATA. Working
for a while for precision physics of simple atoms, which is based
on quantum electrodynamics (QED) calculations, and for fundamental
constants, I have witnessed a certain number of confusions in
applications of the CODATA values. This paper aims to guide to
fundamental constants with a hope to avoid such confusions in
future. One can consider it as a kind of `fundamental constants
for non-experts' or `frequently necessary but not asked
questions'.

Some applications of the values of certain fundamental
constants to precision studies are sensitive to a choice of the values
for the constant to be used. For such a case it is incorrect to
apply any value of the constant blindly. The real option is to
look for the origin of the result, checking what kind of
measurements and calculations have been done to obtain it, what
suggestions were made if any. Before any application of a
particular result on the fundamental constant, one has to realize
whether this application is in line with the actions done to derive
the constant.

The CODATA papers \cite{codata,Mohr2000} represent a very specific
kind of papers, namely, reference papers. They contain very
important information, which can be found on demand, but most of
users are aware only about the tables of the recommended values of
the fundamental constants, and even most of them did not read the
papers, but access to the values through the internet (via, e.g.,
the NIST web site \cite{nistweb}) or through other compilations,
such as the Review of Particle Properties \cite{pdg}. In such a
case they do not even have a chance to see any details of the
original CODATA evaluation.

We consider this note as a supplementary paper to \cite{codata}
and intentionally do not provide any references which can be found
there. We also intentionally do not present any progress since the
adjustment-2002 \cite{codata}. In particular, there have been a
number of remarkable results improving accuracy in determination
of the fine structure constant $\alpha$ and the Planck constant
$h$, as well as substantial progress in understanding the muon
anomalous magnetic moment.

Our purpose is not to discuss the most accurate data for a
particular time period, since the data are continuously improving,
but to explain how to deal with the CODATA recommendations, which
may be applied to any CODATA recommendations, current and future.

Most of physicists consider CODATA as a kind of a brand for
publication of the list of the best values of the constants.
However, the main objective of the CODATA task group on the
fundamental constants is to study the precision data, their
accuracy, reliability and overall consistency. Its papers present
a very detailed critical review of the experimental data which
serve as input data of the adjustments.

\section{The adjustment of the fundamental constants:\\ a general view}

What is the adjustment? Normally, when one performs an experiment, the
final result is an average of various measurements, or a result of
a simple fitting, if we cannot measure the needed values directly,
but only their combinations. For instance, we can measure certain
cross sections as a function of the momentum transfer and the slope
of specifically normalized cross section (as a function of the
momentum transfer squared, $q^2$) gives us a charge radius.

In the case of the fundamental constants the `topology' of
correlation links between data is cumbersome. It may be possible
to measure $e$, $h$, $e/h$, $e^2/h$, $e/m_e$, $h/m_e$ etc. In
contrast to the mentioned scattering experiment, the accuracy of
different results is high, but quite different, and the data
themselves may have also substantial experimental or computational
correlations in uncertainties. The adjustment is such a procedure
which pretends to find the most plausible result for the output
parameters.

It involves a least-square-method as a technical part; however, a
crucial issue is a careful reconsideration of each inconsistency
between and inside various portions of the input data. It depends
on physics whether we have to treat them symmetrically or
asymmetrically. A symmetric treatment may suggest, e.g., either
multiplying their uncertainties by the same factor in order to
reach a reasonable $\chi^2$ value, or, in contrast,
assigning to all the data equal uncertainties despite the fact
they have been claimed to be very different. An example of an asymmetric
treatment is the very removal of certain doubtful data as an
ultimate choice.

\section{The adjustment of the fundamental constants: the data}

All the input data can be subdivided into a few groups as shown in
Table~\ref{t:group} (see, e.g., \cite{physicstoday,ufn} for more
detail). Two `big blocks' involve substantially correlated data of
various kinds (see below). Evaluation of data of these two big
blocks is the main part of the procedure of {\em the adjustment of
the values of the fundamental constants\/}.

Data, which are known with a higher accuracy, can be found
separately before the main adjustment of these two blocks. Those
most accurate data are referred to as {\em auxiliary\/}. An example
of such data is the data on the Rydberg constant $R_\infty$ and
various mass ratios like $m_e/m_p$ (we have to mention also a few
constants such as the speed of light $c$ which numerical
values are fixed in the SI by definition).

Data which are less accurate can be in principle ignored. The
related constants are to be derived afterwards from the results of
the adjustment. An example is a value of $h/(m_ec)$, which is in
principle correlated with a value of the fine structure constant
$\alpha$ (see below); it cannot be directly measured with high
accuracy but can be extracted from adjusted data on $R_\infty$,
$\alpha$ etc. Such data are related to blocks, but only as their
output results.

There are also certain data which are completely {\em
uncorrelated\/} with the two big blocks as, e.g. the results for
the Newtonian constant of gravitation $G$.

\begin{table}[htbp]
\begin{center}
\begin{tabular}{clcc}
\hline
Constant&Value&$u_r$&Comment\\
\hline
$c$&$299\,792\,458\,{\rm m}/{\rm s}$& 0&exact$^*$\\
$\mu_0$&$4\pi \times 10^{-7}\,{\rm N}/{\rm A}^2$& 0&exact$^*$\\
\hline
$R_\infty$&$10\,973\,731.568\,525(73)\;
\mbox{\rm m}^{-1}$ & $[6.6\times10^{-12}]$&auxiliary$^\star$\\
$m_p/m_e$&$1\,836.152\,672\,61(85)$ & $[4.6\times10^{-10}] $&auxiliary$^\star$\\
$m_e$&$5.485\,799\,094\,5(24)\times10^{-4}\;\mbox{\rm u}$ & $[4.4\times10^{-10}]$&auxiliary$^\star$\\
 \hline
$\alpha^{-1}$&$137.035\,999\,11(46)$ & $[3.3\times10^{-9}]$ &$\alpha$-block$^\star$\\
$\lbar_C=\hbar/(m_e c)$&$386.159\,267\,8(26)\times10^{-15}\;\mbox{\rm m}$&$[6.7\times10^{-9}]$&$\alpha$-block$^\dag$\\
$h\, N_A$ &$3.990\,312\,716(27)\times10^{-10}\,\mbox{J\,s/\,mol}^{-1}$&$[6.7\times10^{-9}]$&$\alpha$-block$^\dag$\\
$R_K=h/e^2$ &$25\,812.807\,449(86)\,{\rm \Omega}$& $[3.3\times10^{-9}]$&$\alpha$-block$^\star$\\
\hline
$e$&$1.602\,176\,53(14)\times10^{-19}\,{\rm C}$& $[8.5\times10^{-8}]$&$h$-block$^\ddag$\\
$h$&$6.626\,069\,3(11)\times10^{-34}\,{\rm J}\,{\rm s}$&$[1.7\times10^{-7}]$&$h$-block$^\star$\\
$N_A$&$6.022\,141\,5(10)\times10^{23}\,{\rm mol}^{-1}$&$[1.7\times10^{-7}]$&$h$-block$^\star$\\
$m_e$&$0.510\,998\,918(44)\;{\rm Mev}/c^2 $ & $[8.6\times10^{-8}]$&$h$-block$^\dag$\\
$m_e$&$9.109\,382\,6(16)\times10^{-31}\;\mbox{\rm kg}$ &$[1.7\times10^{-7}]$&$h$-block$^\dag$\\
$K_J=2e/h$&$483\,597.879(41)\times10^{9}\,{\rm Hz}\,{\rm V}^{-1}$&$[8.5\times10^{-8}]$ &$h$-block$^\star$\\
\hline
$G$&$6.674\,2(10)\times10^{-11}\; \mbox{\rm m}^3\mbox{\rm kg}^{-1} \mbox{\rm s}^{-2}  $ & $[1.5\times10^{-4}]$&independent\\
\hline
\end{tabular}
\end{center}
\caption{The recommended values of some fundamental constants
\cite{codata} and their subdivision into the adjustment blocks.
Here, $u_r$ is the relative standard uncertainty. Comments: $^*$
-- fixed by the current definition of the SI units; $^\star$ --
measured and adjusted; $^\dag$ -- derived from the adjusted data;
$^\ddag$ -- $e$ is not measured directly, but its various
combinations with $h$ and $N_A$.} \label{t:group}
\end{table}

The first block is formed by the data related to the fine structure
constant $\alpha$. It also includes the so-called molar Planck
constant $h N_A$ and various results for the particle and atomic
masses in the frequency units (i.e., in the result for the
value $Mc^2/h$ related to the mass $M$). The results in the
frequency units are related to $\alpha$ because of the equation
\begin{eqnarray}
R_\infty &=& \frac{\alpha^2m_ec}{2h}\nonumber\\
&=&\frac{1}{2c}\,\alpha^2\,\frac{Mc^2}{2h}\,\frac{m_e}{M}\;,
\end{eqnarray}
where $M$ is related to the mass of the particle or atom measured
in an experiment in the frequency units and we remind that the
Rydberg constant and a number of important mass ratios $m_e/M$ are
known with higher accuracy.

The molar Planck constant $h N_A$ enters this block as a
conversion factor between two units in which microscopic masses
can be measured with a very high accuracy, namely, the unified
atomic mass units and frequency units.

The other block is formed by somewhat less accurate data related
to the electron charge $e$, the Planck constant $h$ and the
Avogadro constant $N_A$. Because of the high accuracy obtained for
the fine structure constant
\begin{equation}
\alpha=\frac{e^2}{4\pi \epsilon_0 \hbar c}
\end{equation}
and the molar Avogadro constant $h N_A$, the final results for
these three constants are strongly correlated.

\section{Electrical data}

An important feature of these two blocks is a substantial
involvement of electric data related to standards and to some
other macroscopic measurements. Two fundamental constants of
quantum macroscopic effects play an important role there: the von
Klitzing constant
\begin{equation}
R_K = \frac{h}{e^2}=\frac{\mu_0c}{2\alpha}\;,
\end{equation}
which describes the quantized resistance in the quantum Hall
effect, and the Josephson constant
\begin{equation}
K_J =\frac{2e}{h}\;.
\end{equation}

The related data are often refereed in a very confusing way.
For instance, in the so-called measurement of the von Klitzing
constant $R_K$ the crucial part is not the measurement proper, but
a construction of a reference resistance, which should have a
known value in the SI units. The only opportunity for such a
resistance, or rather for an impedance, is based on a so-called
calculable capacitor. Surprising for devices based on classical
physics, the value of the capacitance of certain symmetric
configurations can be set with high accuracy. That is because of
the special topological Thompson-Lampard theorem. Realizations of this
theorem have recently provided us with classical-physics standards of the
SI farad and ohm for a long period. At present a
realization of the Thompson-Lampard capacitor in the only way to
determine a value of $R_K$ directly.

The watt-balance experiments do not involve any balance which deals
with the power. They deal with a special kind of ampere balance
which can be run in the dynamic and static regime. The static regime
involves an electric current, while the dynamic one deals with an
induced potential. Combining two measurements we arrive at a new
quantity, power, as their product with an unknown geometric factor
completely vanishing in the final equation.

A number of electric measurements deal with the gyromagnetic ratio
or the Faraday constant. In practice, they do that in a very
specific way. We have been numerously told from the high school time
that we have to use the International System of Units, the SI,
(despite certain resistance of the physical community). And that
is under control of the International Committee on Weights and
Measures, CIPM. However, the CIPM has sanctioned a departure from
the SI system in precision electric measurements, for which
so-called practical units were recommended in 1990 \cite{cipm}.
The latter, ohm-90 and volt-90, are based on certain fixed values
of $R_K$ and $K_J$ \cite{cipm} and all accurate electric
measurements have been performed in these units.

If one declares a measurement of a certain electric quantity $A$
(e.g., the gyromagnetic ratio of a proton in water), in practice
the value actually measured in the SI units is somewhat more
complicated
\begin{equation}
A R_K^n K_J^m\;,\nonumber
\end{equation}
where $n$ and $m$ are certain integer numbers ($0,\pm1,\pm2$)
which depend on the experiment.

This issue is so non-trivial, that measuring the same quantity,
e.g., the gyromagnetic ratio of a proton, by different methods,
`in a low magnetic field' and `in a high magnetic field', we
arrive at very different results: a determination of $\alpha$ in
former case and of $h$ in the latter, because of difference in
values of $n$ and $m$. That is a kind of a metrological joke because
even the units of the gyromagnetic ratio are different because of
involvement of factors such as $V_{90}/V$. Such factors appear
because in certain situations we cannot avoid applying the SI since
the value of the magnetic constant $\mu_0$ is known exactly in the SI
units and we also have to deal with the practical units as long as
a real measurement is concerned.

One more confusing example is a measurement of the Compton wave
length of a neutron $h/(m_nc)$. The experiment consisted of two
important measurements: one is related to the de Broglie wave
length $\lambda_v=h/(m_nv)$ and the other to the velocity $v$.
They were measured in a sense in quite different units. The
velocity was determined in the proper SI units directly. Meanwhile
the wave length $\lambda_v$ was compared with the lattice spacing
of a certain crystal. This crystal was indirectly compared with a
so-called perfect crystal, basically used for the Avogadro
project. Because of that the $h/m_n$ result is strongly correlated
with a certain block of the data related to $N_A$ and it is not
just an isolated result related to a neutron.

Unfortunately, this customary practice with labelling the results
is very confusing and for a non-expert it is hard to understand
what was really measured and which data are correlated.

\section{Recommended values and the `less accurate' original results}

Now, we can describe the adjustment. In the first approximation,
we have to evaluate the most accurate data only (i.e., the
auxiliary data), next to deal with the results from the $\alpha$ block
and afterwards to adjust the $h$-block. That should give a good
approximate result.

In reality, the less accurate data can still affect more accurate
data, often marginally, but not always. The adjustment is very
similar in a sense to a simple least-square procedure, where the
statistical weight of data drops down with increase of their
uncertainty. However, the less accurate data are still very
important. If they agree with the main part of the data, that
increases the final reliability of the evaluation, which is not
just a question of the $\chi^2$ test. We always want
confirmations, even not very accurate, but independent. However,
with such a large amount of data some may disagree. In such a case
the less accurate data can have very important impact on the
final results.

The data are strongly correlated and one may wonder what should be
done by a user if certain input data are inconsistent as it
actually happens from time to time. If the accuracy of the
application is really sensitive to what value of the constant to
take, one should avoid using the CODATA tables, and use instead
the CODATA analysis of the input data. If accuracy is not
important, it is better to use the same data all over the world,
i.e. the data from the CODATA tables, and it should not matter
whether they are well consistent or not.

\section{The fine structure constant $\alpha$ and related data}

\begin{figure}[hbtp]
\begin{center}
\includegraphics[width=0.7\textwidth]{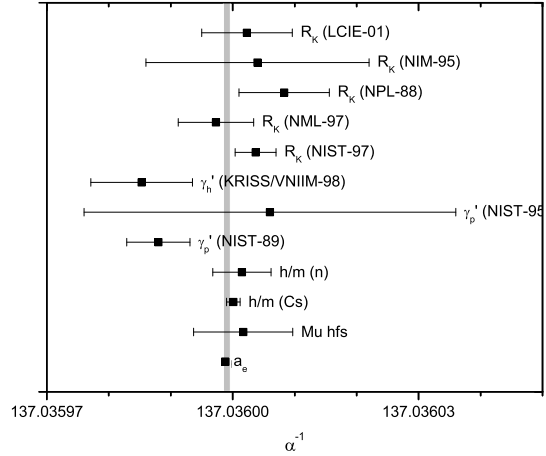}
\end{center}
\caption{The fine structure constant $\alpha$. The vertical strip
is related to the CODATA recommended values. The original results
are explained in \cite{codata}.\label{f:alpha}}
\end{figure}

Let us consider a situation with the fine structure constant as an
example. The CODATA's result
\begin{equation}\label{e:a:co}
\alpha^{-1}=137.035\,999\,11(46)\;,~~~[3.3\times10^{-9}]\;,
\end{equation}
is based mainly on a datum from the anomalous magnetic moment of
an electron $a_e$. All the related contributions are shown in
Fig.~\ref{f:alpha}.

The fine structure constant $\alpha$ plays a crucial role in
quantum electrodynamics (QED) and because of that a few questions
may arise.
\begin{itemize}
\item Could we use the CODATA's value to test QED theory? The
answer is negative. Comparisons of theory and experiment, which
are the most sensitive to a choice of $\alpha$, have already been
included into deduction of the result (\ref{e:a:co}). If we like
to check a particular QED effects we should apply a value of
$\alpha$ obtained without any use of the effects under question.
The CODATA adjusted value includes in principle all QED effects,
for a precision test of which we need an accurate value of $\alpha$.
\item Which value of $\alpha$ can we use then? The answer depends
on what kind of a test we would like to perform. If we would like to test QED
`absolutely', we should take the best non-QED value which is
\begin{equation}\label{e:a:cs}
\alpha^{-1}({\rm
Cs})=137.036\,000\,1(11)\;,~~~[7.7\times10^{-9}]\;,
\end{equation}
a result, derived from the Rahman spectroscopy of the caesium
atom. If we like to check consistency of QED, we can take one of
QED-related values such as
\begin{equation}\label{e:a:g}
\alpha^{-1}(a_e)=137.035\,998\,80(52)\;,~~~[3.8\times10^{-9}]\;,
\end{equation}
and use it for a calculation of other QED effects, such as the
hyperfine interval in the muonium atom.
\item If we calculate a value which is very sensitive to a choice of
$\alpha$ among known values, what have we to do? The best choice
is to reverse the situation, i.e., determine $\alpha$ and
put it into Fig.~\ref{f:alpha}. In this case we can see whether it
agrees with various values. Sometimes the data are not in good
agreement and a new value can completely change the situation.
\item If we like to determine $\alpha$, what is the crucial level
of accuracy? Let us assume for a moment that the data are
perfectly consistent. In such a case the crucial accuracy is that
of the second value in the row, which is (\ref{e:a:cs}). This
value is vital for the reliability of the CODATA result. We remind
that the dominant contribution to (\ref{e:a:co}) comes from the
anomalous magnetic moment and the result (\ref{e:a:g}) has not been
confirmed either experimentally or theoretically.
\item That is not an unusual situation. The most advanced
experiments and calculations are hard to repeat or confirm.
Meanwhile, they have entered `terra incognita' and despite high
quality of the research teams they are most vulnerable because of
lack of experience or rather a wrong `experience' based on trusted
unimportance of various phenomena which may become important. For
instance, for recommendation of conservative committees of CIPM
they sometimes introduce a kind of factor or reliability for
accurate measurements, which may increase the uncertainty tenfold
\cite{meter}.
\item Have we to trust all data for $\alpha$? That is not exactly
the case since there is no appropriate theory for the quantum Hall
effect which provides us with five data points.
\item However, the agreement is good, but not perfect. We note
that two values with the gyromagnetic ratio of a proton are within
certain disagreement with the most accurate value. From a purely
scientific point of view we have a rather good general agreement
(cf. with the situation on $h$ and $G$ below). Nevertheless, there is
an application which deals with a practical unit of resistance by
CIPM \cite{cipm}. They conservatively estimate an uncertainty as a
part in $10^7$. The related value of the fine structure constant
is
\[
\alpha^{-1}({\rm CIPM})=137.035\,997(14)\;,~~~[1\times10^{-7}]\;.
\]
We should mention, however, that CIPM is
overconservative because their results may have legal consequences
and their examinations are for this reason not just a kind of
scientific researches (see \cite{ufn} for further discussion).
\end{itemize}
Actually, that is a strange story how we deal with, e.g.,
$3\sigma$-off points. When they are a part of a large statistics
set of similar measurements, we are satisfied by the $\chi^2$
criterium. Meanwhile, when the data are different such as for the
adjustment of the fundamental constants or QED tests with
different systems, we sometimes pay special attention to such
`bad' points trying to understand what is wrong in their
particular cases.

A comparison of $\alpha$, extracted from a particular QED
value, let us say, the muonium hyperfine interval, after a certain
improvement of theory, with other $\alpha$'s has a number of
additional reasons (in respect to a comparison with the CODATA
recommended value only).
\begin{itemize}
\item The muonium datum has been already used for determination of
$\alpha$ in (\ref{e:a:co}). Despite the fact that is has a
marginal effect, it is not appropriate to compare a certain
improvement of $\alpha({\rm Mu hfs})$ with an average value, which
includes an earlier version of $\alpha({\rm Mu hfs})$. The new and
old values are based on the same experiment and the very
appearance of the new value means that the old value is out of
date.
\item If we have a contradiction, we can clearly see whether the
new value contradicts to one or two most accurate data but agrees
with the most of the rest or so, or it disagrees with all.
\item Known data experience corrections from time to time. Using
a set of original data, one can introduce the proper corrections.
However, there is no way to correct the CODATA value, except
indeed redoing the adjustment.
\end{itemize}

The latter is a result of a complicated procedure which includes
re-examination of accuracy of various data and test of their
accuracy. It is not possible to update the list of recommended
values very often. Because of that a substantial delay may take
place. For instance, the most recent CODATA paper was published in
2005 and we can expect a new one in 2008. The deadline for the
input data in \cite{codata} was the end of 2002. That means that
any evaluation including data obtained since 2003 will not be
available until 2008. Because of that it may be important in
certain cases to consider original results reviewed in the recent
CODATA paper \cite{codata} and add there new results, available
since recently, if any.

\section{The Planck constant $h$ and related data}

\begin{figure}[hbtp]
\begin{center}
\includegraphics[width=0.7\textwidth]{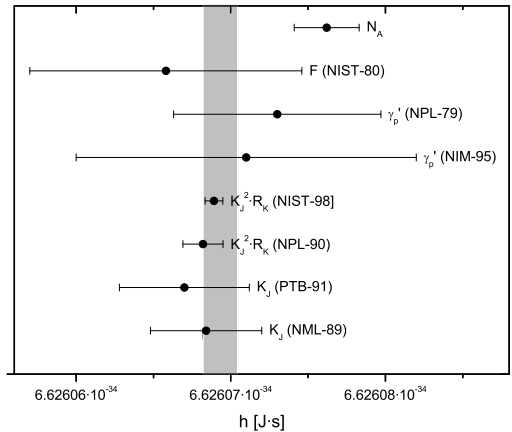}
\end{center}
\caption{The Planck constant. The vertical strip is related to the
CODATA recommended values. The original results are explained in
\cite{codata}.\label{f:planck}}
\end{figure}

Determination of the fine structure constant has demonstrated
a rather good agreement. The situation is not always so good.
As an important example of a substantially worse agreement we
present data related to the Planck constant $h$ in
Fig.~\ref{f:planck}.

The data are not in good agreement. In particular, a result
related to $N_A$ contradicts to the most accurate data obtained
from the watt-balance experiments. We will return to this result
later. We need to mention that CIPM recommended a value of the
Josephson constant $K_J=2e/h$ with a conservative uncertainty of 2
parts in $10^7$ while their conservative value of $R_K=h/e^2$ has
uncertainty of a part in $10^7$. The related value for the Planck
constant is
\[
h({\rm CIPM})=6.626\,068\,9(53)\times10^{-34}~{\rm J}\,{\rm
s}\;,~~~[8.1\times10^{-7}]\;.
\]

\section{The Newtonian constant of gravitation}

\begin{figure}[hbtp]
\begin{center}
\includegraphics[width=0.6\textwidth]{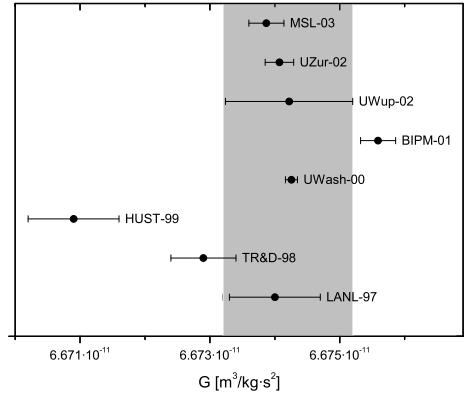}
\end{center}
\caption{The Newtonian constant of gravitation. The vertical strip
is related to the CODATA recommended values. The original results
are explained in \cite{codata}.\label{f:bigg}}
\end{figure}

The results on the Newtonian constant of gravitation $G$ show an
even much worse situation with a scatter superseding the
uncertainty by many times (see Fig.\ref{f:bigg}).

Despite the gravitation constant is without any doubt one of the most
fundamental constants, its accuracy does not have great
importance. Fundamentality of $G$ shows itself first of all in the
application to quantum gravity where the obtained results are
rather qualitative than quantitative. Another important application
is due to general relativity. Precision tests of general
relativity involve much higher accuracy than the one in the
determination of the Newtonian constant. For actual problems, the
most important constant is a product of a gravitating mass (of Sun
or Earth) and $G$ and such products have been known much more
accurately than $G$ and from completely different kind of data.

Still there is a kind of experiments of fundamental nature which are
in part similar to measurements of $G$, namely, studies of
equivalence principle in laboratory distance scale. However, such
experiments are differential and essential part of uncertainty
should cancel out.

As a result, we note that the determination of $G$ is indeed an
ambitious and important problem, but it is somewhat separated from
both the rest of the precision data and applications of
fundamental physics.

\section{The fundamental constants and their numerical values}

The discussion above raises a more general question on fundamental
constants and their values. The numerical value of a dimensional
fundamental constant involves the units and thus involves a certain
kind of phenomena which are used to determine units. Such an
involvement can change the physical meaning when going from the
constant to its value drastically.

While the constants, such as the speed of light or the Planck
constant are determined by Nature, their numerical values can be
treated with a certain room for arbitrariness. We can, e.g.,
adopt certain numerical values by definition.

In the case of variability of the constants, the interpretation of
possible changes of the constants and their numerical values is
quite different (see, e.g., \cite{looking}).

Two constants, discussed above, $h$ and $G$, are truly
fundamental, but they are not very often needed for accurate
calculations. Below we consider certain values more closely related
to atomic and particle physics or, in more general terms, to
microscopic physics.

\section{Microscopic and macroscopic quantities}

In microscopic physics nobody intends to apply any macroscopic unit
such as a kilogram. However, the nature of the units is not a
trivial issue. We should distinguish between their rough values
and their definition. Rough values of various units have been
determined historically. For most of the SI units they are
macroscopic, such as for a kilogram or a second. Meanwhile, the
SI kilogram is defined as a macroscopic unit, but the SI second at
present is defined as a kind of atomic unit via the hyperfine
interval in caesium-133 atom.

The only SI unit which has a clear historic microscopic sense is
the volt. To proceed with potentials one dealt with breaking
atomic or molecular bonds. A characteristic ionization potential
is of a few volts and, in particular, in hydrogen it is about
13.6~V. A popular non-SI unit, the electron-volt possesses in a
rough consideration a clear atomic sense. Because of this
ionization issue, an energy, related to $R_\infty$ is, indeed,
13.6 eV. However, if we look at the definition of the volt in a
practical way, we find that the volt of the SI is defined via the
ampere and the watt. The latter are defined via the kilogram, the
metre, the second and a fixed value of the magnetic constant of
vacuum $\mu_0$. Because of presence of the kilogram, the volt and
the electron-volt have macroscopic meaning from the point of view
of measurements.

Measuring microscopic values in terms of macroscopic units is
always a complicated problem, which introduces serious unnecessary
uncertainties. Meanwhile, the very use of the electron-volt in the
atomic, nuclear and particle physics is an issue completely based
on a custom and never related to real matter. It is a kind of
illusion. However, for missing a difference between reality and
illusion, one has to pay. The price is an unnecessary uncertainty
in various data, expressed in the electron-volts and a correlation
between uncertainties of various data.

The electron-volt is widely used in microscopic physics. In
particular, it is customarily applied to characterize the X-ray and
gamma-ray transitions by their energy and to present particle
masses in units of GeV/c$^2$. We have to emphasize that nobody
performs any precision measurement in these units in practice. The
transitions are measured in relative units. To measure them
absolutely one has to apply X-ray optical interferometry and
either compare an X-ray and an optical wave length, or
calibrate a lattice parameter in a certain crystal in terms of an
optical wave length. That means that in actual precision
measurements one really deals with the wave length (or related
frequency) and not with the energy of the transition.

The most accurate relative measurements of hard radiation are in
fact more accurate than the conversion factor between the
frequency and the energy, namely, $e/h$ (if the energy is measured
in the electron-volts). The uncertainty of this coefficient is
presently $8.5\times10^{-8}$ \cite{codata}. We strongly recommend
for transition frequencies measured more accurately than 1~ppm
to present results in frequency units and for results in the
electron-volts to present separately two uncertainties: of the
measurement and of the conversion into electron-volts. It
would be also helpful to specify explicitly the value of the
conversion factor used.

If one even tries to measure energy in electron-volts `by
definition', the electron-volts proper are still not the best
choice. CIPM recommended a practical unit, volt-90, in terms of
which the Josephson constant $K_J$ has an exactly fixed value
\cite{cipm}. In such a case, the result would be expressed in
terms of eV$_{90}$, rather than in eV's. The uncertainty of the
conversion factor $e/h$ in practical units is zero.

Mev's and Gev's are also widely used for the masses of particles
and for the energy excess in nuclear physics. From the point of view
of accuracy, such units are not better than kilograms. The best
choice is to apply direct results of relative measurements (mass
ratios), when available, or to express the masses in terms of
either of the two adequate microscopic units. One of the latter is the
unified atomic mass unit, u, and the other corresponds to the
frequency related to $mc^2/h$. In these two units elementary
masses are known with the highest accuracy.

\section{Reliability of the input data and the recommended values}

The easiest part of the evaluation is their mutual evaluation. Two
most important questions are related to the data.

1. Not all available data are included as input data and not all
input data are exactly equal to the originally published data. The
question to decide prior to the evaluation is how to treat each piece
of data? Should we accept them "as they are", or assign them a
corrected uncertainty, or even dismiss some of them prior to any
evaluation procedure? That should be decided on base of quality of
the data.

2. After initial probe mutual least-square evaluations are done, we
used to see that some pieces are not in perfect agreement with
the rest of the data. That cannot be avoided once we have many
pieces of data. That opens another important question, to be decided
at the initial stage of the evaluation. How should we treat the data
when they are combined together? In other words, should we do
anything with the data due to their inconsistency if any? At this
stage the decision is partly based on their consistency, partly on
their correlations and partly still on their initial properties.

These questions are to be decided not on base of statistics (like
when in an easy case of a number of data points for the same
quantity one drops the smallest and the largest results) but first
of all on base of their origin, their experimental and theoretical
background.

The CODATA's recommended values are the best one, but in principle
that does not mean all of them are really good. They are the best
because the authors perform the best possible evaluation of
existing data. If data are not good enough, the result of any
evaluation cannot be good. The CODATA task group are not
magicians. That is why it is essential to have independent
results for each important quantity. Below we consider a question of
the reliability of data important in atomic and particle physics.

The conservative policy of CIPM and discrepancy in the input data
(see Fig.~\ref{f:planck}) show that direct use of the CODATA
result is not a single option to be considered. The CIPM treatment of
the data does not contradict to the CODATA approach, because CIPM
applies the CODATA analysis; however, prefers to derive a more
conservative result from the CODATA's consideration.

\begin{figure}[hbtp]
\begin{center}
\includegraphics[width=0.6\textwidth]{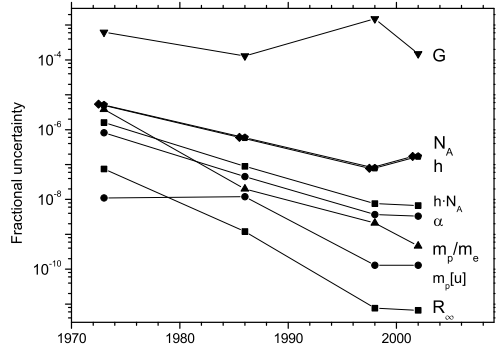}
\end{center}
\caption{Progress in determination of fundamental constants by the
CODATA task group (see \cite{codata,Mohr2000} and references
to earlier results therein).\label{f:prog}}
\end{figure}

An important illustration of reliability of the recommended values
is presented in Fig.~\ref{f:prog}. While for most of them progress
with time reduced the uncertainty, sometimes (e.g., for $h$ or
$G$) better understanding meant appearance of a discrepancy.

\section{Proton properties}

Among the particles listed in CODATA tables \cite{codata} two, a proton
and muon, are of particular interest.
 The most confusing datum
on a proton is its charge radius, $R_p$. The CODATA paper recommends the
result
\begin{equation}
R_p ({\rm CODATA)}= 0.8750(68)\;{\rm fm}\;,
\end{equation}
which, in principle, is based on all available data including
electron scattering and hydrogen spectroscopy. Nevertheless, we
would not recommend to apply this result blindly to any sensitive
issue. The dominant contribution comes from spectroscopy of
hydrogen and deuterium and the related theory. The spectroscopic
data included various experiments, which partly confirm each
other. However, a substantial progress made in the theory (the
Lamb shift) is related to surprisingly large higher-order two-loop
terms \cite{Pachucki}, which are neither understood qualitatively
nor independently confirmed quantitatively. I would not consider the
theoretical expressions at the moment as a reliable result until their
proper confirmation or understanding. Such a need for an
independent confirmation is a characteristic issue for any
breakthrough in either theory or experiment.

The second (in terms of accuracy) result mentioned in
\cite{codata}
\begin{equation}
R_p ({\rm Sick)}= 0.895(18)\;{\rm fm}\;,
\end{equation}
is the one obtained by Sick \cite{sick} from the examination of world
scattering data. This piece of CODATA input data is very specific.
CODATA very seldom accepts any evaluation of world data without
performing a critical reconsideration. A crucial feature of the
CODATA treatment of the world data is reconsideration of accuracy
on experimental and theoretical results. The most important
scattering results were obtained long time ago. They dealt with
QED scattering corrections obtained a few decades ago. At present,
the QED corrections are known better, but there is no simple way
to reevaluate the existing scattering data. The Sick's
examination is the most competent I have ever seen. But it is an
evaluation of the data ``as they were published''.

The problem of correcting the experimental data because of a
possibly unappropriate treatment of higher-order radiative
corrections by the original authors, was not addressed in his
evaluations. I would rather consider the central value of this
evaluation as a valid one but would somewhat increase the
uncertainty (see discussion in \cite{report}) achieving
\begin{equation}
R_p = 0.895(30)\;{\rm fm}\;.
\end{equation}

It is hard to be more precise with the uncertainty. If such a
reevaluation were done in the CODATA paper, the problem should be
addressed. But in was not done in \cite{codata}. A reason not to
do that is twofold.

First, it is an obvious fear that the job could not be done
properly because of lack of necessary information for experiments
done long time ago. Next, the proton size from the scattering
plays rather minor role in the adjustment. The evaluation of the
auxiliary block with the Rydberg constant is sensitive to theory
of a so-called state-dependent part of the Lamb shift of the $n$
states and to theory of states with a non-zero orbital moment.
Both depend on a value of the proton size marginally. That means
that CODATA evaluation of the Rydberg constant needs only a very
rough value of the proton size and we can accept any result for
$R_p$ for such an evaluation.

The recommended value of the proton charge radius is
actually determined by the same spectroscopic study. The rest of
the data can rather produce a marginal effect on the value of $R_p$.
In particular, the second value of the radius, obtained from the
scattering, is rather out of interest of the CODATA evaluation and
they do not care about it. The reevaluation of the world
scattering data from the CODATA side looks like an unnecessary
overcomplicated problem with unclear reliability of the outcome.

One more proton property of interest is its magnetic moment, or
rather electron-to-proton ratio of the magnetic moments
\begin{equation}
\frac{\mu_e}{\mu_p} = 658.210\,6860(66)\;,~~~[1\times10^{-8}]\;.
\end{equation}
The result is completely based on an MIT experiment performed long
time ago \cite{mit}. While for the most important constants such as
$\alpha$ and $h$ one can easily find all sources
for particular results in \cite{codata}, it is hard to see what
result is the second in accuracy. While details of the analysis will be
published elsewhere, here we conclude that the data may be obtained from
a study of the muonium magnetic moment and the most accurate partial
result
\begin{equation}
\frac{\mu_e}{\mu_p} = 658.210\,70(15)\;,~~~[2.3\cdot10^{-7}]
\end{equation}
is much less accurate than the MIT value.

\section{Muon properties}

The muon data include the muon magnetic moment, mass and $a_\mu$,
the anomalous magnetic moment. The latter should not be used at all
for any sensitive issue. The CODATA can make a reasonable
prediction only after the situation is settled, while for $a_\mu$
it is not. Speaking more generally, CODATA is a brand for
the best constants, but not all products with this brand are
equally good. Critical examination of input data can improve their
reliability and reduce their scatter. I would say that is the most
competent evaluation of world data on the fundamental constants.
Nevertheless, there is no magics in the CODATA adjustment and the
result cannot be better than the input data allow. Before trusting
any particular CODATA result one has to take a look into the
data analysis.

The result for $a_\mu$ has contributions from experiment and
theoretical evaluations based on $e^+e^-$ and $\tau$ data. To
consider physics we should not average these partial results but
reexamine and compare them.

The mass and magnetic moment have been used numerously in a quite
confusing way. The experiment, most sensitive to their values,
has been included into the evaluation. For instance, one can
apply a value of $m_e/m_\mu$ (or $\mu_\mu/\mu_p$) to the hyperfine
interval in muonium, either assuming QED to determine $\alpha$, or
accepting a certain value of $\alpha$ to verify QED. However, the
CODATA results
\begin{eqnarray}
\frac{m_\mu}{m_e}&=&206.768\,283\,8(54)\;,~~~[2.6\times10^{-8}]\;,\nonumber\\
\frac{\mu_\mu}{\mu_p}&=&3.183\,345\,118(89)\;,~~~[2.6\times10^{-8}]\;,
\end{eqnarray}
are dominated by a value extracted from the muonium hyperfine
interval assuming a certain value of $\alpha$ and validity of QED.
The second best set
\begin{eqnarray}
\frac{m_\mu}{m_e}&=&206.768\,276(24)\;,~~~[1.2\times10^{-7}]\;,\nonumber\\
\frac{\mu_\mu}{\mu_p}&=&3.183\,345\,24(37)\;,~~~[1.2\times10^{-7}]\;,
\end{eqnarray}
comes from separate data and may be used to either determine
$\alpha$ or test QED.

We emphasize that all this information is contained in the CODATA
papers \cite{Mohr2000,codata}; however, since `simple users' are
more interested just in the tables they usually miss it.

We remind that there is a number of compilations of various kinds
of data around the world and even reading carefully most of
the compilations, there is no chance to find detail of input data.
Sentences such as `the uncertainty does not include systematic
error' or so are often missing when a datum came from the original
paper to a compilation. The CODATA paper is one of very few
exceptions, however, a way of reader's treatment of the CODATA
papers sometimes doesn't make use of this advantage.

\section{Impact of a redefinition of the kilogram on values of the
fundamental constants}

To conclude the paper, we would like to discuss two issues. One is rather
technical and related to a possible redefinition of the kilogram
and the ampere in terms of fixed values of $h$ and $e$
\cite{definition}. It is most likely that this redefinition will
be adopted, but it is unclear when. The numerical values of the
fundamental constants play two roles. One is that they represent
in a numerical way certain experimental data. Redefining the
kilogram, obviously the experimental results would not change and
the information would not be added. Still, a certain pieces of the
information and related uncertainty can be removed from some data.
After redefinition of units, certain experiments done with a
relatively low accuracy could be isolated from the fundamental
constants (e.g., any direct study of the prototype of the kilogram
would have no relation to basic physical quantities anymore). The
other role of the numerical values is that they are reference
data. As we mentioned above, it is customary to use, without any
experimental or theoretical reasons, the electron-volts. The
redefinition of the kilogram and the ampere would establish them
as microscopic units (and the volt as well). The conversion factor
$e/h$ would be known exactly. That means that all values in
electron-volts would have adequate accuracy.

\section{Legacy of the adjustment of the fundamental constant}

The last question to discuss here is a conceptual one. Doing
precision physics, we cannot ignore the very fact that we accept a
large number of physical laws. Sometimes they are proved with a
certain accuracy, sometimes they are not.

For instance, there is no accepted theory which demands that the
electron charge and the proton charge be of the same value. We
have various direct experimental tests, but those are always
limited by their accuracy. The conceptual evidence should come from
a new theory, which is confirmed experimentally. We strongly expect
a certain unification theory, but no evidence has been available
up-to-date.

We expect that the fundamental constants are really constant, but we
do not understand their origin and we (or most of us) believe that
during the inflation epoch of the universe some constants such as
$m_e/m_p$ changed. So, the constancy of the constants is merely an
experimental fact. What is even more important, certain
physical laws are put into the very base of our system of units,
the SI, and if they would occur incorrect, one may wonder whether
that is detectable. The answer is positive. If we adopt a set of
assumptions, either with an internal inconsistency or inconsistent
with Nature, we should be able to see either an inconsistency in
the interpretation of the results (e.g., a contradiction within two
determinations of the same quantity) or a discrepancy between the
trusted assumption and the observed reality.

To test any particular law, one has to rely on specific experiments
sensitive to such a violation. The CODATA examination is mainly
based on the assumption that we can follow the known physical
laws. We know that any particular physical theory is an
approximation. Combining the data from different fields we check
the consistency of the overall picture (both: the laws and the
approximations) and the result obtained is satisfactory. Up to
now.

I am grateful to Simon Eidelman for useful discussions. This work
was supported in part by the RFBR grant \# 06-02-16156 and DFG
grant GZ 436 RUS 113/769/0.

\end{document}